\documentclass[letter,11pt]{article}
\usepackage{etex}
\usepackage[top=1in,bottom=1in,left=1in,right=1in]{geometry}
\usepackage{slashed}
\usepackage{epsfig}
\usepackage{comment}
\usepackage{cancel}
\usepackage{bbm}
\usepackage{array}
\usepackage{bigints}
\usepackage{booktabs}
\usepackage{color}
\usepackage{dsfont}
\usepackage{float}
\usepackage{framed}
\usepackage{graphicx}
\usepackage{indentfirst}
\usepackage{mathrsfs}
\usepackage{multirow} 
\usepackage{subdepth}
\usepackage{subfig}
\usepackage{titlesec}
\usepackage[dotinlabels]{titletoc}
\usepackage{wrapfig}
\usepackage[all]{xy}
\usepackage{young}
\usepackage[vcentermath]{youngtab}
\usepackage{relsize}
\usepackage{hyperref}
\usepackage{feynmp}
\usepackage{appendix}
\numberwithin{equation}{section}

\usepackage[utf8]{inputenc}
\usepackage{geometry}
\usepackage{graphicx}
\usepackage{array}
\usepackage{subfig}
\usepackage{slashed}
\usepackage{amsmath}
\usepackage{amsfonts}
\usepackage{amssymb}
\usepackage{cite}
\usepackage{epsfig}
\usepackage{comment}
\usepackage{tikz}
\usepackage{feynmp}
\usepackage{cancel}
\usepackage{mathrsfs}
\usepackage{mathtools}

\usepackage{cleveref}
\crefname{section}{§}{§§}
\Crefname{section}{§}{§§}

 \def\p{\partial}
 \def\bz{{\bar z}}

\def\0{{(0)}}
\def\1{{(1)}}
\def\2{{(2)}}

\def\<{\langle }
\def\>{\rangle }

\newcommand{\bea}{\begin{eqnarray}}
\newcommand{\eea}{\end{eqnarray}}
\newcommand{\be}{\begin{equation}}
\newcommand{\ee}{\end{equation}}
\newcommand{\ba}{\begin{align}}
\newcommand{\ea}{\end{align}}

   \makeatletter
  \let\over=\@@over \let\overwithdelims=\@@overwithdelims
  \let\atop=\@@atop \let\atopwithdelims=\@@atopwithdelims
  \let\above=\@@above \let\abovewithdelims=\@@abovewithdelims
\renewcommand\section{\@startsection {section}{1}{\z@}%
                                   {-3.5ex \@plus -1ex \@minus -.2ex}%nn
                                   {2.3ex \@plus.2ex}%
                                   {\normalfont\large\bfseries}}

\renewcommand\subsection{\@startsection{subsection}{2}{\z@}%
                                     {-3.25ex\@plus -1ex \@minus -.2ex}%
                                     {1.5ex \@plus .2ex}%
                                     {\normalfont\bfseries}}

\newcommand{\beq}{\begin{equation}}
\newcommand{\eeq}{\end{equation}}
\newcommand{\beqa}{\begin{eqnarray}}
\newcommand{\eeqa}{\end{eqnarray}}
\newcommand{\beqar}{\begin{eqnarray*}}

\def\[{\big[}
\def\]{\big]}

\def\bz{{\bar z}}

\def\be{{\bar \epsilon}}

\def\ca{{\mathcal A}}

\def\eps{\epsilon}

\def\be{\begin{equation}}
\def\ee{\end{equation}}

\def\+{{(+)}}
\def\-{{(-)}}
\def\0{{(0)}}
\def\1{{(1)}}
\def\2{{(2)}}
\def\3{{(3)}}

\newcommand{\bd}[1]{\begin{fmffile}{#1}\begin{fmfgraph*}}
\newcommand{\ed}{\end{fmfgraph*}\end{fmffile}}

\begin{document}
\begin{titlepage}
\unitlength = 1mm
\ \\
\vskip 3cm
\begin{center}

{\LARGE{\textsc{Conformally Soft Theorem in Gauge Theory}}}

\vspace{0.8cm}
Monica Pate, Ana-Maria Raclariu and Andrew Strominger

\vspace{1cm}

{\it  Center for the Fundamental Laws of Nature, Harvard University,\\
Cambridge, MA 02138, USA}

\vspace{0.8cm}

\begin{abstract}

Asymptotic particle states in four-dimensional celestial scattering amplitudes are 
labelled by their $SL(2,\mathbb{C})$ Lorentz/conformal weights $(h,\bar h)$ rather than the usual energy-momentum four-vector. These boost eigenstates involve a superposition of all energies. As such, celestial 
gluon (or photon) scattering cannot obey  the usual (energetically) soft theorems. In this paper we show that tree-level celestial 
gluon scattering, in theories with sufficiently soft UV behavior,  instead obeys {\it conformally soft } theorems involving $h\to 0$ or $\bar h\to 0$. 
Unlike the energetically soft theorem, the conformally soft theorem cannot be derived from low-energy effective field theory.

 \end{abstract}

\vspace{1.0cm}

\end{center}

\end{titlepage}

\pagestyle{empty}
\pagestyle{plain}

\def\vx{{\vec x}}
\def\p{\partial}
\def\po{$\cal P_O$}

\pagenumbering{arabic}
 
%%%%%%%%%%%%%%%%---------------------END OF TITLE PAGE AND ABSTRACT---------------------%%%%%%%%%%%%%%%%%%%%%%

\tableofcontents

\section{Introduction}
  Four-dimensional (4D) Minkowskian scattering amplitudes are usually presented in a momentum space basis, in which translations act trivially by phases, while the $SL(2,\mathbb{C})$ action of the Lorentz group is more nontrivial.  The latter acts as the conformal group on the celestial two-sphere at null infinity. Recently there has been interest in and explicit construction of amplitudes in a conformal basis on the celestial sphere \cite{He:2015zea,Cheung:2016iub,Pasterski:2016qvg,Pasterski:2017kqt,Cardona:2017keg,Strominger:2017zoo,Pasterski:2017ylz,Lam:2017ofc,Banerjee:2017jeg,Schreiber:2017jsr,Banerjee:2018gce,Stieberger:2018edy, Donnay:2018neh,Stieberger:2018onx,Banerjee:2019aoy,Fan:2019emx}, in which the conformal action is trivial but translation invariance is hidden.  We will use the symbol $\widetilde \ca$  to denote such amplitudes (and distinguish them from the usual amplitudes denoted $\ca$) and simply refer to them as celestial  amplitudes. The structure of celestial amplitudes has so far turned out to be surprisingly rich. Properties which are obscure  in $\ca$ are sometimes obvious in $\widetilde \ca$ and vice versa.   Celestial amplitudes are of interest for many reasons, including a holographic representation of quantum gravity in flat space \cite{deBoer:2003vf,Barnich:2009se,Barnich:2010eb,Barnich:2011ct,Kapec:2014opa, He:2015zea,Kapec:2016jld,Cheung:2016iub,Strominger:2017zoo,Kapec:2017gsg}.  
  
A central feature of the usual scattering amplitudes $\ca$ in gravity or gauge theory is the existence of a variety of (energetically) soft theorems. These greatly constrain, and in some cases completely determine \cite{Arkani-Hamed:2017jhn,Rodina:2018pcb}, all  amplitudes. 
On the other hand, for celestial amplitudes there is no energy which can be taken to be soft. Instead of the energy, external particle states 
are characterized by their $SL(2,\mathbb{C})$ conformal dimensions $(h, \bar h)$, which are roughly their  boost weight plus or minus spin. 
This raises the question: \vspace{5pt}\\
{\it Is there an analog of the soft theorem which constrains celestial amplitudes? } \vspace{5pt}\\
The main result of this paper is to answer this question in the affirmative for nonabelian gauge theories at tree level. The usual energetically soft theorem is replaced by 
the {\it conformally soft theorem}, in which the left or right conformal dimension ($h$ or $\bar h$) of an external particle is taken to zero. The demonstration uses the energetically soft relations of $\ca$ and is valid only when the UV behavior is sufficiently soft. Overlapping observations have been made in a number of papers including \cite{Cheung:2016iub, Fan:2019emx, Schreiber}. 

From some points of view this result is expected and obvious. However, on further inspection,  our  derivation encounters  some subtleties and relies crucially on assumed good UV behavior of the momentum space amplitudes $\ca$. Moreover it does not fit into our usual effective field theory  worldview. The energetically soft theorem is understood to be valid because very soft photons or gluons cannot  probe the shorter-distance structure of the scattering region, and can sense only the external particle trajectories which extend to infinity. Conformally soft particles, on the other hand,  involve arbitrarily high energies, and their scattering might have been expected to involve UV microphysics. The fact that this is {\it not} the case cannot be understood from low-energy reasoning, but seems natural from the celestial vantage point. 

The constraints on $\ca$ imposed by the various  conformally soft theorems strongly resemble those of a current algebra on a 2D conformal field theory, while for example  supertranslation invariance (not discussed herein) provides an infinite number of a fundamentally new type of constraints which recursively relate operators with different dimension \cite{Strominger:2013jfa,Donnay:2018neh,Stieberger:2018onx}.  One would like to combine all known  constraints and construct the most general compatible scattering amplitude. One example of an  interesting question, which  has been raised for example in \cite{Arkani-Hamed:2017jhn,Rodina:2018pcb}, is \vspace{5pt}\\
{\it How does one characterize the most general 4D four-graviton scattering amplitude consistent with all known constraints?} \vspace{5pt}\\
The celestial picture provides a new framework for this question. Perhaps the answer is not impossibly far away. 

In this paper we consider only tree-level amplitudes. 
Of course the energetically soft theorem is generically corrected at one-loop level in nonabelian gauge  theory. It would be very interesting to understand the nature of the loop  corrections  to the conformally soft theorem, whose form is highly constrained by the exact  conformal symmetry of celestial amplitudes. See \cite{Campiglia:2019wxe} for recent  progress on  celestial loop corrections. 

This paper is organized as follows. In section 2 we provide a general argument for the conformally soft theorem which applies to amplitudes with sufficiently soft UV behavior. This motivates a conjecture on the existence of perturbative quantum deformations of tree-level amplitudes in four-dimensional quantum field theory, which we present in section 3. The conformally soft theorem is explicitly verified in section 4 for the known tree-level celestial amplitudes, including 4-point Type I and Heterotic string amplitudes as well as $n$-point MHV amplitudes in Yang-Mills theory. 

\section{Conformally soft from soft}
In this section we derive a general identity for the scattering of conformally soft gluons in gauge theory. 
The main inputs are the (energetically) soft gluon theorem and assumed good UV behavior of momentum space amplitudes. 

	 The identity is expressed in terms of ``celestial amplitudes'', denoted  $\widetilde \ca$, of gluons in a conformal primary basis  given by the Mellin transform of the standard momentum space amplitudes $\ca$ \cite{Pasterski:2017kqt, Pasterski:2017ylz}
	\be
	\label{mag}
		\widetilde \ca_{J_1 \cdots J_n} (\lambda_i, z_i, \bz_i) =\left( \prod_{k = 1}^n \int_0^\infty d \omega_k ~ \omega_k^{i \lambda_k} \right)\ca_{J_1 \cdots J_n} (\omega_i, z_i, \bz_i),
	\ee
	where $J_i$ label both the $4$D and $2$D helicities.  We parametrize a null four-momentum by a sign $\eps_i = \pm 1$ (for outgoing and ingoing particles respectively), a positive frequency $\omega_i$ and a point on the celestial sphere $(z_i, \bz_i)$ such that 
	\be \label{parmom}
		p_i^\mu = \eps_i \omega_i q^\mu(z_i, \bz_i),
	\ee
	where
	\be
q^{\mu}(z,\bz) =  \Big(1 + z\bz, z  + \bz, -i(z - \bz), 1 - z\bz\Big).
\ee
	The $\lambda_i$ are related to the conformal dimensions $\Delta_i =1 + i \lambda_i$ which in turn are related to $(h, \bar{h})$ by $(h, \bar{h}) = \frac{1}{2}(\Delta+J, \Delta -J)$. One may show \cite{Pasterski:2017kqt, Pasterski:2017ylz} that the celestial amplitudes have the  standard $SL(2,\mathbb{C})$ Lorentz transformation property
	\be 
	\widetilde{\mathcal{A}}_{J_1\cdots J_n}\left(\lambda_i, \frac{a z_i + b}{cz_i + d}, \frac{\bar{a}\bz_i + \bar{b}}{\bar{c}\bz_i + \bar{d}}\right) = \prod_{j=1}^n \left[\left(c z_j + d \right)^{\Delta_j + J_j}\left(\bar{c}\bz_j + \bar{d} \right)^{\Delta_j - J_j}\right] \widetilde{\mathcal{A}}_{J_1\cdots J_n}(\lambda_i, z_i, \bz_i).
\ee
 The momentum space amplitudes are conventionally normalized with inner product\footnote{We use here the conventions of  \cite{ Pasterski:2017ylz}. Helicities of all particles are labelled relative to the outgoing direction from the interaction. }
	 \be
	(p_1,J_1; p_2,J_2) =  (2\pi)^32  p_1^0   \delta^{(3)}(\vec{p}_1 + \vec{p}_2) \delta_{J_1, -J_2}.
	 \ee
This implies the celestial inner product
	\be
	(\lambda_1, z_1, \bz_1, J_1; \lambda_2, z_2, \bz_2, J_2) = (2\pi)^4 \delta(\lambda_1 + \lambda_2) \delta^{(2)}(z_1 - z_2) \delta_{J_1, - J_2}.
	\ee

	Let us consider the $\lambda_j \to 0$ limit, which for a positive (negative) helicity gluon corresponds to taking $\bar{h}$ ($h$) to zero
	\be \label{lim1}
		\begin{split}
			\lim_{\lambda_j \to 0} i \lambda_j \widetilde \ca_{J_1 \cdots J_n} (\lambda_i, z_i, \bz_i)
			& = \Big(\prod_{\substack{k = 1\\ k\neq j}}^n \int_0^\infty d \omega_k ~ \omega_k^{i \lambda_k} \Big)
			 	\left( \int_0^\infty d \omega_j  \lim_{\lambda_j \to 0}  i \lambda_j\omega_j^{i \lambda_j-1 } \right)  \omega_j \ca_{J_1 \cdots J_n} (\omega_i, z_i, \bz_i)\\
			& = 2 \Big( \prod_{\substack{k = 1\\ k\neq j}}^n \int_0^\infty d \omega_k ~ \omega_k^{i \lambda_k} \Big)
			 	   \int_0^\infty d \omega_j  \delta (\omega_j) \omega_j \ca_{J_1 \cdots J_n} (\omega_i, z_i, \bz_i).
		\end{split}
	\ee

	Importantly,  the second equality is valid only under the assumption that the momentum space amplitudes vanish sufficiently rapidly at large momenta. This is because 
 we have used the  representation of the Dirac delta function  
	\be \label{deltaid1}
		\delta(x) = \lim_{\eps \to 0} \frac{\eps}{2} |x|^{\eps-1},
	\ee
which  is only valid when integrated against functions which fall off at least as fast as $x^{-b}$ for $b>\eps$ as $x \rightarrow \infty$.\footnote{To see this, note the integral 
		\begin{equation*}
			\int_0^\infty dx~ \eps x^{\eps -1} \frac{1}{(x+a)^b} = \frac{1}{a^{b-\eps}} \frac{\eps \Gamma ( b-\eps) \Gamma (\eps)}{\Gamma (b)}, \quad \quad  a, b >0,
		\end{equation*} where $b > {\rm Re}(\eps)$ in order for the integral to converge at the upper limit.} To study the behavior of the integrand in \eqref{lim1}, we change variables $\omega_k \rightarrow \omega_k \omega_j$ for every $k \neq j$ yielding 	\be 
		\begin{split}
			 \widetilde \ca_{J_1 \cdots J_n} (\lambda_i, z_i, \bz_i) &=\bigg( \prod_{\substack{k = 1\\ k\neq j}}^n \int_0^\infty d \omega_k ~ \omega_k^{i \lambda_k} \bigg)
			 	\int_0^{\infty}d\omega_j ~\omega_j^{i\sum_k\lambda_k -1} \omega_j^{n} \ca_{J_1 \cdots J_n} ( \omega _j \omega_1, z_1, \bz_1; ...;\omega_j ,z_j,\bz_j;...) .
		 \end{split}
	\ee
Hence we can use the representation of the delta function \eqref{deltaid1} provided
	\be
	\label{lim}
\lim_{\omega_j \rightarrow \infty}	 \omega_j^{n} \ca_{J_1 \cdots J_n} ( \omega _j \omega_1, z_1, \bz_1; ...;\omega_j ,z_j,\bz_j;...) = \mathcal{O}\left( \omega_j^{-\eps}\right).
	\ee  Tree-level  pure Yang-Mills amplitudes scale homogeneously under the rescaling of all momenta
	\be
		\omega_j^{n} \ca^{\text{Yang-Mills}}_{J_1 \cdots J_n} ( \omega _j \omega_1, z_1, \bz_1; ...;\omega_j ,z_j,\bz_j;...)  = 
			\ca^{\text{Yang-Mills}}_{J_1 \cdots J_n} (   \omega_1, z_1, \bz_1; ...;1,z_j,\bz_j;...),
	\ee
	and hence marginally  $violate$  \eqref{lim}.   However, in string theory the tree-level four-gluon amplitudes are exponentially suppressed at large momenta \cite{Veneziano:1968yb}
	\be
		\lim_{\omega_j \rightarrow \infty} \omega_j^{4} \ca^{{\rm string}}_{J_1 \cdots J_4} ( \omega _j \omega_1, z_1, \bz_1; ...;\omega_j ,z_j,\bz_j;...) \sim e^{- \alpha' \omega_j^2},
	\ee
	so the use of this representation of the delta function is justified in this case.  
		
Our approach will be to  $\it define$ the pure Yang-Mills celestial amplitudes as the $\alpha' \rightarrow 0$ limit of string amplitudes. 
More generally various types of UV and other singularities are encountered in celestial amplitudes at both tree and loop level.
This approach potentially provides a systematic regulator for all of these singularities.

	In the theory so defined, we find that the $\lambda_j \rightarrow 0$ limit selects the residue of the pole in the momentum space amplitude at $\omega_j =0$, which is determined by the soft theorem.  Using the (energetically) soft theorem for color-ordered partial amplitudes  
	\be
	\label{msst}
		\lim_{\omega_j \to 0} \omega_j \ca_{J_1 \cdots J_n} (\omega_i, z_i, \bz_i) =- \frac{1}{2} \frac{z_{j-1~j+1}}{z_{j-1~j} z_{j~j+1}} \ca_{J_1 \cdots J_{j-1} J_{j+1} \cdots J_n} (\omega_i, z_i, \bz_i),
	\ee
	we obtain the tree-level  conformally soft theorem\footnote{ In \cite{Fan:2019emx}, the authors present a different normalization of the amplitudes which  removes the poles in $\lambda_j$.  However, for $n \geq 4$, that choice of  normalization yields celestial amplitudes
	which  are not invariant under the translation symmetry discussed in  \cite{Stieberger:2018onx}. The basis chosen here preserves manifest translation invariance.}
	\be \label{MellinWard}
		\lim_{\lambda_j \to 0} i \lambda_j  \widetilde \ca_{J_1 \cdots J_n} (\lambda_i, z_i, \bz_i)
			 = -\frac{1}{2} \frac{z_{j-1~j+1}}{z_{j-1~j} z_{j~j+1}}\widetilde \ca_{J_1 \cdots J_{j-1} J_{j+1} \cdots J_n} (\lambda_i, z_i, \bz_i) .
	\ee

\section{A conjecture}

As we have just seen, the transformation from standard scattering amplitudes (in a momentum basis) to 
celestial amplitudes (in a conformal basis) involves an integral over all energies. Hence, the mere existence of the celestial amplitudes requires soft  UV behavior even at tree level.  In Yang-Mills theory, the UV behavior is just marginally soft enough. In the scalar theories studied in  \cite{Lam:2017ofc,Pasterski:2016qvg,Cardona:2017keg}, the celestial amplitudes are also finite. In Einstein gravity, on the other hand, the 
Mellin transforms diverge and even the classical four-graviton celestial amplitudes do not exist, while they do appear to exist in classical string theory. It is intriguing that soft  UV behavior is required even for the existence of the classical amplitudes. Of course it is also known that there are a variety of  restrictions on the UV behavior of classical scattering (such as the Froissart-Martin bound) required for a unitary quantum perturbation theory. 
Could these be the same restrictions? This motivates the following \vspace{5pt}\\
 {\sc Conjecture: }{\it  A perturbative quantum deformation of  a classical four-dimensional Minkowskian quantum field theory exists if and only if the celestial amplitudes $\widetilde \ca_{J_1...J_n}(\lambda_i,z_i,\bz_i)$ exist.}\vspace{5pt}\\
This conjecture is consistent with the (rather small) number of examples which have been analyzed so far. 
\section{Explicit examples}
As both a consistency check on our formal argument of section 2 and to gain familiarity with celestial structures, in this section we verify that the conformally soft theorem holds for the known examples of tree-level gluon scattering in Yang-Mills and string theory. 
\subsection{Review of gluon amplitudes }
	\label{rev:cp34}
In this subsection we review the expressions for  celestial gluon amplitudes in both Yang-Mills and string theory \cite{Pasterski:2017ylz, Schreiber:2017jsr, Stieberger:2018edy}.

	The general formula for $n$-point MHV Yang-Mills amplitudes in momentum space is
	\be \label{momentumMHV}
		\begin{split}
		\ca_{--+\cdots +}(p_1,..., p_n) &= \frac{\langle 12 \rangle ^3}{\langle 23 \rangle \cdots \langle n1 \rangle} \delta ^{(4)}\Big (\sum_{i = 1}^n p_i\Big)\\
			& = 
			\frac{1}{(-2)^{n-4}} \frac{z_{12}^3}{z_{23} \cdots z_{n1}}\frac{\omega_1 \omega_2}{\omega_3 \cdots \omega_n} \delta ^{(4)}\Big (\sum_{i = 1}^n \eps_i \omega_i q_i\Big),
		\end{split}
	\ee
	where in the last equality, we used $p_i=\eps_i \omega_i q_i$ as in  \eqref{parmom} and
	\be \label{anglebrackdef}
		\langle ij \rangle = -2 \eps_i \eps_j \sqrt{\omega_i \omega_j} z_{ij} .
	\ee
	Here and hereafter powers of the gauge (or string) coupling are absorbed into the wave function normalization. 
To study the $3$-point amplitude, we work in $(2,2)$ signature, where the amplitude is non-vanishing. $z_i$ and $\bz_i$ are then independent real variables, enabling us to rewrite the momentum-conserving delta function as
	\be
		\begin{split}
		\label{3pt-mc}
			\delta^{(4)} \Big(\sum_{i = 1}^3 \eps_i \omega_i q_i \Big) &=\frac{{\rm sgn}(z_{23} z_{31})}{4\omega_3^2 z_{23}z_{31}} \delta \Big( \omega_1- \frac{\omega_3  \eps_3 }{\eps_1} \frac{z_{23} }{z_{12} } \Big)\
				\delta \Big(\omega_2 - \frac{\omega_3  \eps_3}{\eps_2} \frac{ z_{31}}{z_{12} } \Big)\delta  ( \bz_{13} )\delta ( \bz_{23} ),
		\end{split} 
	\ee
	where the form above assumes $z_{ij} \neq 0.$ An alternate form may be found by instead assuming $\bz_{ij} \neq 0$, 
	\be 
	\delta^{(4)} \Big(\sum_{i = 1}^3 \eps_i \omega_i q_i \Big) = \frac{{\rm sgn}(\bz_{23} \bz_{31})}{4\omega_3^2 \bz_{23}\bz_{31}} \delta \Big( \omega_1- \frac{\omega_3  \eps_3 }{\eps_1} \frac{\bz_{23} }{\bz_{12} } \Big)\
				\delta \Big(\omega_2 - \frac{\omega_3  \eps_3}{\eps_2} \frac{ \bz_{31}}{\bz_{12} } \Big)\delta  ( z_{13} )\delta ( z_{23} ).
	\ee
For $--+$ scattering we use \eqref{3pt-mc} in which case the resulting celestial amplitude is
	\be \label{3ptfinalb} 
		\begin{split}
			\widetilde \ca_{--+}(\lambda_i, z_i, \bz_i)  
				& =-  \pi \delta (\lambda_1 + \lambda_2 + \lambda_3) 
					 \frac{    z_{12}^3}{  z_{23}^2  z_{31}^2 } \Big(  \frac{\eps_3 }{\eps_1} \frac{z_{23} }{z_{12} } \Big)^{1+ i \lambda_1} \Big(  \frac{  \eps_3}{\eps_2} \frac{ z_{31}}{z_{12} } \Big)^{1+ i \lambda_2}
				 	\\& \quad \quad  \times 
					 {\rm sgn}(z_{23} z_{31})\delta  ( \bz_{13} )\delta ( \bz_{23} )  
						 \Theta \Big(   \frac{ \eps_3 }{\eps_1} \frac{z_{23} }{z_{12} } \Big)\
			\Theta \Big(  \frac{   \eps_3}{\eps_2} \frac{ z_{31}}{z_{12} } \Big).
		\end{split}
	\ee
Here $\Theta$ vanishes (equals unity) for negative (positive) argument. 
	
	Similarly, to determine the celestial 4-point amplitude, the trick is to rewrite the momentum-conserving delta function
	\be \label{4pointdeltafinal405}
		\begin{split}
			\delta^{(4)} \Big( \sum_{i = 1}^4 \eps_i \omega_i q_i\Big)
				& = \frac{1}{4 \omega_3} \delta \Big(\omega_1-\frac{\omega_3 \eps_3 }{ \eps_1}\frac{   z_{23}  \bz_{34} }{z_{12}\bz_{14} } \Big)
				 	 \delta \Big( \omega_2-\frac{\omega_3 \eps_3 }{ \eps_2} \frac{  z_{13} \bz_{34}  }{z_{12} \bz_{42}}\Big)\\& \quad \quad \times 
					 \delta \Big(\omega_4- \frac{\omega_3 \eps_3 }{ \eps_4}\frac{\bz_{13} z_{23}  }{\bz_{14} z_{42}  } \Big) \delta \Big(z_{12} z_{34} \bz_{13} \bz_{24}-z_{13} z_{24} \bz_{12} \bz_{34} \Big) ,
		\end{split}
	\ee
	from which it directly follows that 
	\be \label{Mellin4pointYang-Mills}
		\begin{split}
			\widetilde \ca&_{--++} (\lambda_i, z_i , \bz_i)\\
				& = \frac{\pi}{2}  \delta \Big(\sum_{i = 1}^4 \lambda_i\Big)  \frac{z_{12}^3}{z_{23} z_{34} z_{41}}  \Big(\frac{\eps_3}{\eps_1}\frac{   z_{23}  \bz_{34} }{z_{12}\bz_{14} } \Big)^{1+ i \lambda_1}
					 \Big(\frac{\eps_3}{\eps_2}\frac{  z_{13} \bz_{34}  }{z_{12} \bz_{42}}\Big)^{1+i  \lambda_2}
					\Big(\frac{\eps_3}{\eps_4}\frac{\bz_{13} z_{23}  }{\bz_{14} z_{42}  } \Big)^{-1+i  \lambda_4} \\& \quad \quad \times  
						 \delta \Big(z_{12} z_{34} \bz_{13} \bz_{24}-z_{13} z_{24} \bz_{12} \bz_{34} \Big)  
					  \Theta\Big( \frac{  \eps_3 }{ \eps_1}\frac{   z_{23}  \bz_{34} }{z_{12}\bz_{14} } \Big)
				 	 \Theta \Big(  \frac{ \eps_3 }{ \eps_2} \frac{  z_{13} \bz_{34}  }{z_{12} \bz_{42}}\Big) 
					 \Theta \Big(  \frac{  \eps_3 }{ \eps_4}\frac{\bz_{13} z_{23}  }{\bz_{14} z_{42}  } \Big)  .
		\end{split}
	\ee
	The factor of $\delta \left(\sum \lambda_i\right)$ arises from the scale invariance of Yang-Mills amplitudes in momentum space, and generalizes to the $n$-point amplitude. 
	
	In momentum space, the 4-gluon amplitudes in Type I and Heterotic string theory differ from the Yang-Mills amplitude by a multiplicative factor \cite{Green:1981xx,Gross:1985rr}
	\be
		\begin{split}
			\ca_{--++}^{\rm Type\ I} = \frac{\Gamma(1-   s) \Gamma(1-  u)}{\Gamma(1- s-  u)} \ca_{--++}^{\text{Yang-Mills}} , \quad
			\ca_{--++}^{\rm Heterotic} = - \frac{\Gamma(-   s)\Gamma(-   t)\Gamma(-  u)}{\Gamma(  s)\Gamma(  t)\Gamma(  u)}  \ca_{--++}^{\text{Yang-Mills}},
		\end{split}
	\ee
	where $s = p_{12}, \ u = p_{23}, \ t = p_{13}$ with
	\be 
	p_{ij} = -\alpha'( p_i + p_j)^2 = 4\alpha' \epsilon_i \epsilon_j \omega_i \omega_j z_{ij}\bz_{ij}.
	\ee
As string amplitudes are not scale invariant, the $\delta \left(\sum \lambda_i\right)$ factor no longer appears. 	As shown  in \cite{Stieberger:2018edy}, the celestial 4-gluon amplitudes in Type I and Heterotic string theory 
are obtained by the replacements  
	\be \label{defFI}
		(2\pi) \delta \Big(\sum_{i = 1}^4 \lambda_i\Big) \to \int_{0}^{\infty} d \omega ~ \omega^{i ( \lambda_1 + \lambda_2 + \lambda_3 + \lambda_4 ) -1 } 
			\left( \frac{\Gamma(1-   s(\omega)) \Gamma(1-  u(\omega))}{\Gamma(1- s(\omega)-  u(\omega))}\right) \equiv F_{I}
	\ee
	and
	\be \label{defFH}
		(2\pi) \delta \Big(\sum_{i = 1}^4 \lambda_i\Big) \to \int_{0}^{\infty} d \omega ~ \omega^{i ( \lambda_1 + \lambda_2 + \lambda_3 + \lambda_4 ) -1 } 
			\left (- \frac{\Gamma(-   s(\omega))\Gamma(-   t(\omega))\Gamma(-  u(\omega))}{\Gamma(  s(\omega))\Gamma(  t(\omega))\Gamma(  u(\omega))} \right)\equiv F_H
	\ee
	in \eqref{Mellin4pointYang-Mills}, respectively. Here
	\be \label{defMandomega}
		\begin{split}
			  s(\omega) =  4 \alpha' \omega^2\frac{   z_{23}  \bz_{13} }{z_{42} \bz_{14} } z_{34} \bz_{34}, \ \
			  u(\omega) =  4 \alpha' \omega^2  \frac{  z_{13} \bz_{34}  }{z_{12} \bz_{42}}  z_{23} \bz_{23}, \ \
			  t(\omega) =  4 \alpha' \omega^2 \frac{   z_{23}  \bz_{34} }{z_{12}\bz_{14} }  z_{13} \bz_{13}.
		\end{split}
	\ee 
	
A salient feature of celestial string amplitudes with five or more gluons is - assuming soft high-energy behavior - that they appear to be finite at generic values of $(\lambda_i, z_i)$. This 
	contrasts with the momentum-space amplitudes, which vanish when momentum is not conserved and are infinite 
	when it is (due to the delta-function).  For four gluons,  however, all celestial amplitudes develop a kinematic delta-function singularity 
	arising from translation invariance, which restricts the 2D conformal cross-ratio to lie on the real axis. This can be seen directly from \eqref{4pointdeltafinal405}.  For three gluons further kinematic singularities appear in $(2,2)$ signature, while in 
	$(3,1)$ the amplitudes vanish.  In the $\alpha'\to 0$ Yang-Mills limit, the celestial amplitudes are no longer generically finite due to the appearance of the $\delta\left(\sum \lambda_i\right)$ factor. While the regulation and properties of momentum space delta-function singularities is long-understood, the delta-function singularities of celestial amplitudes are rather different and remain to be fully understood.

	\subsection{4-point  Yang-Mills amplitude}
	
		In this section we explicitly study the conformally soft limit of the 4-point Yang-Mills gluon amplitude \eqref{Mellin4pointYang-Mills}.  Taking $\lambda_4 \to 0$ and using the identity \eqref{deltaid1} introduces an additional delta function on the 
		right-hand side of  \eqref{Mellin4pointYang-Mills}
		\be   \label{4pointintA}
		\begin{split}
			\lim_{\lambda_4 \to 0}& i \lambda_4 \widetilde \ca_{--++} (\lambda_i, z_i , \bz_i)\\
				& = \delta \Big(\frac{\eps_3}{\eps_4}\frac{\bz_{13} z_{23}  }{\bz_{14} z_{42}  } \Big) \Bigg[ \pi \delta \Big(\sum_{i = 1}^3 \lambda_i\Big)  \frac{z_{12}^3}{z_{23} z_{34} z_{41}}  \Big(\frac{\eps_3}{\eps_1}\frac{   z_{23}  \bz_{34} }{z_{12}\bz_{14} } \Big)^{1+ i \lambda_1}
					 \Big(\frac{\eps_3}{\eps_2}\frac{  z_{13} \bz_{34}  }{z_{12} \bz_{42}}\Big)^{1+i  \lambda_2}
					 \\& \quad \quad \times  
						 \delta \Big(z_{12} z_{34} \bz_{13} \bz_{24}-z_{13} z_{24} \bz_{12} \bz_{34} \Big)  
					  \Theta\Big( \frac{  \eps_3 }{ \eps_1}\frac{   z_{23}  \bz_{34} }{z_{12}\bz_{14} } \Big)
				 	 \Theta \Big(  \frac{ \eps_3 }{ \eps_2} \frac{  z_{13} \bz_{34}  }{z_{12} \bz_{42}}\Big) 
					 \Theta \Big(  \frac{  \eps_3 }{ \eps_4}\frac{\bz_{13} z_{23}  }{\bz_{14} z_{42}  } \Big) \Bigg] .
		\end{split}
	\ee
	Assuming 
	\be
		z_{23} \neq0,
	\ee
	where we are taking $z_i$ and $\bz_i$ to be independent real variables, the additional delta function can be written as
	\be \label{deltasimp1}
		\delta \Big(\frac{\eps_3}{\eps_4}\frac{\bz_{13} z_{23}  }{\bz_{14} z_{42}  } \Big) ={\rm sgn}(\bz_{14} z_{42} z_{23}) \frac{\bz_{14} z_{42} }{ z_{23} } \delta (\bz_{13}). 
	\ee
	Then, \eqref{4pointintA} simplifies to 
	\be   \label{4pointintB}
		\begin{split}
			\lim_{\lambda_4 \to 0}& i \lambda_4 \widetilde \ca_{--++} (\lambda_i, z_i , \bz_i)\\
				& = {\rm sgn}(\bz_{14} z_{42} z_{23}) \frac{\bz_{14} z_{42} }{ z_{23} } \delta (\bz_{13})
					 \Bigg[ \frac{1}{4} (2\pi) \delta \Big(\sum_{i = 1}^3 \lambda_i\Big) 
					  \frac{z_{12}^3}{z_{23} z_{34} z_{41}}
					   \Big(\frac{\eps_3}{\eps_1}\frac{   z_{23}    }{z_{12}  } \Big)^{1+ i \lambda_1}
					     \\& \quad \quad \times  
					 \Big(\frac{\eps_3}{\eps_2}\frac{  z_{13} \bz_{34}  }{z_{12} \bz_{42}}\Big)^{1+i  \lambda_2} 
						 \delta ( z_{13} z_{24} \bz_{12} \bz_{34}  )  
					  \Theta\Big( \frac{  \eps_3 }{ \eps_1}\frac{   z_{23}   }{z_{12}  } \Big)
				 	 \Theta \Big(  \frac{ \eps_3 }{ \eps_2} \frac{  z_{13} \bz_{34}  }{z_{12} \bz_{42}}\Big) 
					\Bigg] ,
		\end{split}
	\ee
	where we find the argument of one of the step functions vanishes and we take $\Theta (0 ) = \frac{1}{2}$.
	
	To further simplify, we focus on the other delta function by assuming\footnote{We make such assumptions here because in this section we are interested in the conformally soft singularity structure of the amplitude
	as opposed to the collinear singularity structure. We do not study the  regime in which the singularities overlap.}
	\be
		z_{13} \neq 0,  \quad  \quad z_{24} \neq 0 \quad {\rm and } \quad \bz_{34} \neq 0, 
	\ee
so that
	\be \label{deltasimp2}
		\delta ( z_{13} z_{24} \bz_{12} \bz_{34}  )   =\frac{ {\rm sgn}(z_{13} z_{24}   \bz_{34} ) }{z_{13} z_{24}   \bz_{34}} \delta( \bz_{12}).
	\ee	Using this, the amplitude \eqref{4pointintB} further simplifies 
	\be   \label{4pointintC}
		\begin{split}
			\lim_{\lambda_4 \to 0}& i \lambda_4 \widetilde \ca_{--++} (\lambda_i, z_i , \bz_i)\\
				& =   \frac{ {\rm sgn}( z_{23}z_{31})  }{z_{31}    z_{23} } \delta( \bz_{12})\delta (\bz_{13}) 
					 \Bigg[ \frac{1}{4} (2\pi) \delta \Big(\sum_{i = 1}^3 \lambda_i\Big) 
					  \frac{z_{12}^3}{z_{23} z_{34} z_{41}}
					   \Big(\frac{\eps_3}{\eps_1}\frac{   z_{23}    }{z_{12}  } \Big)^{1+ i \lambda_1}
					 \Big(\frac{\eps_3}{\eps_2}\frac{  z_{31}    }{z_{12}  }\Big)^{1+i  \lambda_2} 
					     \\& \quad \quad \times  
					  \Theta\Big( \frac{  \eps_3 }{ \eps_1}\frac{   z_{23}   }{z_{12}  } \Big)
				 	 \Theta \Big(  \frac{ \eps_3 }{ \eps_2} \frac{  z_{31}   }{z_{12}  }\Big) 
					 \Bigg] .
		\end{split}
	\ee
	Comparing with our expression for the 3-point celestial amplitude \eqref{3ptfinalb}, we verify the anticipated relation \eqref{MellinWard} 
	\be   \label{4pointYang-MillsWard}
		\begin{split}
			\lim_{\lambda_4 \to 0}&  i \lambda_4  \widetilde \ca_{--++} (\lambda_i, z_i , \bz_i) 
				 = -\frac{1}{2} \frac{z_{31}}{z_{34} z_{41}}\widetilde \ca_{--+}(\lambda_i, z_i, \bz_i)  .
		\end{split}
	\ee
	
	\subsection{4-point Type I amplitude}
	
		In this subsection  we study the conformally soft limit of the 4-point Type I gluon amplitude
		\be \label{Mellin4pointTypeI}
		\begin{split}
			\widetilde \ca&_{--++}^{\rm Type~I} (\lambda_i, z_i , \bz_i)\\
				& = \frac{1}{4}F_I \frac{z_{12}^3}{z_{23} z_{34} z_{41}}  \Big(\frac{\eps_3}{\eps_1}\frac{   z_{23}  \bz_{34} }{z_{12}\bz_{14} } \Big)^{1+ i \lambda_1}
					 \Big(\frac{\eps_3}{\eps_2}\frac{  z_{13} \bz_{34}  }{z_{12} \bz_{42}}\Big)^{1+i  \lambda_2}
					\Big(\frac{\eps_3}{\eps_4}\frac{\bz_{13} z_{23}  }{\bz_{14} z_{42}  } \Big)^{-1+i  \lambda_4} \\& \quad \quad \times  
						 \delta \Big(z_{12} z_{34} \bz_{13} \bz_{24}-z_{13} z_{24} \bz_{12} \bz_{34} \Big)  
					  \Theta\Big( \frac{  \eps_3 }{ \eps_1}\frac{   z_{23}  \bz_{34} }{z_{12}\bz_{14} } \Big)
				 	 \Theta \Big(  \frac{ \eps_3 }{ \eps_2} \frac{  z_{13} \bz_{34}  }{z_{12} \bz_{42}}\Big) 
					 \Theta \Big(  \frac{  \eps_3 }{ \eps_4}\frac{\bz_{13} z_{23}  }{\bz_{14} z_{42}  } \Big)  ,
			\end{split}
		\ee
		with $F_I$ given in \eqref{defFI}.  Performing the same limit as in pure Yang-Mills \eqref{4pointintA}, again we obtain  an additional delta function on the 
		right-hand side of  \eqref{Mellin4pointTypeI} 
		\be   \label{4pointintA_typeI}
		\begin{split}
			\lim_{\lambda_4 \to 0}& i \lambda_4 \widetilde \ca_{--++}^{\rm Type~I} (\lambda_i, z_i , \bz_i)\\
				& = \delta \Big(\frac{\eps_3}{\eps_4}\frac{\bz_{13} z_{23}  }{\bz_{14} z_{42}  } \Big) \Bigg[ \frac{1}{2} \left( \left. F_I \right|_{\lambda_4 = 0}\right)  \frac{z_{12}^3}{z_{23} z_{34} z_{41}}  \Big(\frac{\eps_3}{\eps_1}\frac{   z_{23}  \bz_{34} }{z_{12}\bz_{14} } \Big)^{1+ i \lambda_1}
					 \Big(\frac{\eps_3}{\eps_2}\frac{  z_{13} \bz_{34}  }{z_{12} \bz_{42}}\Big)^{1+i  \lambda_2}
					 \\& \quad \quad \times  
						 \delta \Big(z_{12} z_{34} \bz_{13} \bz_{24}-z_{13} z_{24} \bz_{12} \bz_{34} \Big)  
					  \Theta\Big( \frac{  \eps_3 }{ \eps_1}\frac{   z_{23}  \bz_{34} }{z_{12}\bz_{14} } \Big)
				 	 \Theta \Big(  \frac{ \eps_3 }{ \eps_2} \frac{  z_{13} \bz_{34}  }{z_{12} \bz_{42}}\Big) 
					 \Theta \Big(  \frac{  \eps_3 }{ \eps_4}\frac{\bz_{13} z_{23}  }{\bz_{14} z_{42}  } \Big) \Bigg] .
		\end{split}
	\ee	Assuming as before that $z_{23} \neq 0$ so \eqref{deltasimp1} holds, we find the right-hand side of \eqref{4pointintA_typeI} reduces to the same expression as in pure Yang-Mills \eqref{4pointintB}, since
	from \eqref{defFI} and \eqref{defMandomega}, one readily finds
	\be
		  F_{I} \Big|_{\substack{\lambda_4 = 0\\ \bz_{13} = 0}}
			 = \int_{0}^{\infty} d \omega ~ \omega^{i ( \lambda_1 + \lambda_2 + \lambda_3  ) -1 }  = 2\pi \delta( \lambda_1 + \lambda_2 + \lambda_3  ) .
	\ee
	Then, by following the remaining steps that were applied to the pure Yang-Mills amplitude, one concludes
	\be   \label{4pointTypeIWard}
		\begin{split}
			\lim_{\lambda_4 \to 0}& i \lambda_4  \widetilde \ca_{--++}^{\rm Type~I} (\lambda_i, z_i , \bz_i) 
				 = -\frac{1}{2} \frac{z_{31}}{z_{34} z_{41}}\widetilde \ca_{--+}(\lambda_i, z_i, \bz_i) .
		\end{split}
	\ee
	The  amplitude appearing on  the right-hand side is the 3-point celestial amplitude \eqref{3ptfinalb}, which is identical in Yang-Mills and Type I string theory. 
	
	\subsection{4-point  Heterotic amplitude}
	
		From the previous section, we learn that in order to show that the 4-point amplitude in Heterotic string theory obeys the same Ward identity as the 4-point Type I and Yang-Mills amplitudes, we first need
		to determine 
		\be
			 F_{H} \Big|_{\substack{\lambda_4 = 0\\ \bz_{13} = 0}} = 
			 	\int_{0}^{\infty} d \omega ~ \omega^{i ( \lambda_1 + \lambda_2 + \lambda_3   ) -1 } 
			\left (- \frac{ \Gamma(-  u(\omega))}{ \Gamma(  u(\omega))} \right).
		\ee
	 	This readily follows from the equations \eqref{defFH} and \eqref{defMandomega} and the fact
		\be
			\lim_{x \to 0} \frac{\Gamma(x)}{\Gamma(-x)} = -1. 
		\ee
	 	Using this and following the steps outlined in the previous section, we arrive at
		\be   \label{4pointintBhet}
		\begin{split}
			\lim_{\lambda_4 \to 0}& i \lambda_4 \widetilde \ca_{--++}^{\rm Heterotic} (\lambda_i, z_i , \bz_i)\\
				& = {\rm sgn}(\bz_{14} z_{42} z_{23}) \frac{\bz_{14} z_{42} }{ z_{23} } \delta (\bz_{13})
					 \Bigg[ \frac{1}{2} \Big(F_{H} \Big|_{\substack{\lambda_4 = 0\\ \bz_{13} = 0}} \Big)
					  \frac{z_{12}^3}{z_{23} z_{34} z_{41}}
					   \Big(\frac{\eps_3}{\eps_1}\frac{   z_{23}    }{z_{12}  } \Big)^{1+ i \lambda_1}
					     \\& \quad \quad \times  
					 \Big(\frac{\eps_3}{\eps_2}\frac{  z_{13} \bz_{34}  }{z_{12} \bz_{42}}\Big)^{1+i  \lambda_2} 
						 \delta ( z_{13} z_{24} \bz_{12} \bz_{34}  )  
					  \Theta\Big( \frac{  \eps_3 }{ \eps_1}\frac{   z_{23}   }{z_{12}  } \Big)
				 	 \Theta \Big(  \frac{ \eps_3 }{ \eps_2} \frac{  z_{13} \bz_{34}  }{z_{12} \bz_{42}}\Big) 
					 \Theta (0) \Bigg] .
		\end{split}
	\ee
	 Assuming a non-collinear configuration such that \eqref{deltasimp2} applies,  the right-hand side of \eqref{4pointintBhet} reduces to the 
	 expression for pure Yang-Mills \eqref{4pointintC} since
	 \be
			 F_{H} \Big|_{\substack{\lambda_4 = 0\\ \bz_{13} = 0\\ \bz_{12}= 0}} =(2 \pi) \delta (  \lambda_1 + \lambda_2 + \lambda_3   )   .
	\ee
	Thus, the Heterotic 4-gluon amplitude likewise obeys the Ward identity 
	\be   \label{4pointhetWard}
		\begin{split}
			\lim_{\lambda_4 \to 0}& i \lambda_4  \widetilde \ca_{--++}^{\rm Heterotic} (\lambda_i, z_i , \bz_i) 
				 = -\frac{1}{2} \frac{z_{31}}{z_{34} z_{41}}\widetilde \ca_{--+}(\lambda_i, z_i, \bz_i).
		\end{split}
	\ee
	
\subsection{Permuting external legs}
	In momentum space, the four-point MHV amplitude is symmetric under the exchange of 3 and 4 and consequently the soft limits $\omega_3$ and $\omega_4$ can both be readily obtained from the momentum-space
	expression \eqref{momentumMHV}.  On the other hand, the four-point celestial amplitudes were written in subsection \ref{rev:cp34}  in a form that elucidates their behavior in the $\lambda_4 \to 0$ conformally soft limit, 
	but obscures their behavior in the $\lambda_3 \to 0$ conformally  soft limit.  Of course one can  start from  the momentum-space amplitudes and  derive a different  expression for the celestial amplitudes
	 in which the $\lambda_3 \to 0$ conformally  soft limit can be easily  taken. However, one of the goals of this paper is to understand everything directly in the conformal basis, so  in this subsection we explain how to permute 3 and 4  without  reference  to the momentum-space representation. 
We shall see the permutation symmetry is related to scaling properties of the amplitudes. 	 
	 The general form of the 4-gluon celestial  amplitude is
	 \be  
		\begin{split}
			\widetilde \ca&_{--++} (\lambda_i, z_i , \bz_i)\\
				& = \frac{1}{4}  F(\lambda_i, \hat s, \hat u)  \Big(\frac{\eps_3}{\eps_1}\frac{   z_{23}  \bz_{34} }{z_{12}\bz_{14} } \Big)^{1+ i \lambda_1}
					 \Big(\frac{\eps_3}{\eps_2}\frac{  z_{13} \bz_{34}  }{z_{12} \bz_{42}}\Big)^{1+i  \lambda_2}
					\Big(\frac{\eps_3}{\eps_4}\frac{\bz_{13} z_{23}  }{\bz_{14} z_{42}  } \Big)^{-1+i  \lambda_4} \\& \quad \times  
						\frac{z_{12}^3}{z_{23} z_{34} z_{41}}  \delta \big(z_{12} z_{34} \bz_{13} \bz_{24}-z_{13} z_{24} \bz_{12} \bz_{34} \big)  
					  \Theta\Big( \frac{  \eps_3 }{ \eps_1}\frac{   z_{23}  \bz_{34} }{z_{12}\bz_{14} } \Big)
				 	 \Theta \Big(  \frac{ \eps_3 }{ \eps_2} \frac{  z_{13} \bz_{34}  }{z_{12} \bz_{42}}\Big) 
					 \Theta \Big(  \frac{  \eps_3 }{ \eps_4}\frac{\bz_{13} z_{23}  }{\bz_{14} z_{42}  } \Big)  ,
		\end{split}
	\ee
	where 
	\be
		F(\lambda_i, \hat s, \hat u) = \int_0^\infty d\omega ~ \omega^{-1 + i\sum_{i = 1}^4 \lambda_i } f(\omega^2 \hat s,\omega^2 \hat u)
	\ee
	and $\omega^2 \hat s = s$ and $\omega^2 \hat u = u$, with $s$ and $u$ given in \eqref{defMandomega} and related to $t$ by $s+t+ u = 0$.
	
	Now suppose we perform a change of variables $\omega \to \omega \alpha$, where $\alpha>0$.  We find
	\be
		F(\lambda_i, \hat s, \hat u)  = \alpha^{ i\sum_{i = 1}^4 \lambda_i } F(\lambda_i, \alpha^2\hat s, \alpha^2\hat u) 
	\ee
	and in the full amplitude, 
	 \be  
		\begin{split}
			\widetilde \ca&_{--++} (\lambda_i, z_i , \bz_i)\\
				& = \frac{1}{4}  F(\lambda_i,\alpha^2 \hat s, \alpha^2\hat u)  \Big(\alpha\frac{\eps_3}{\eps_1}\frac{   z_{23}  \bz_{34} }{z_{12}\bz_{14} } \Big)^{1+ i \lambda_1}
					 \Big(\alpha\frac{\eps_3}{\eps_2}\frac{  z_{13} \bz_{34}  }{z_{12} \bz_{42}}\Big)^{1+i  \lambda_2}
					\Big(\alpha\frac{\eps_3}{\eps_4}\frac{\bz_{13} z_{23}  }{\bz_{14} z_{42}  } \Big)^{-1+i  \lambda_4} \alpha^{-1 + i \lambda_3} \\& \quad \times  
						\frac{z_{12}^3}{z_{23} z_{34} z_{41}}  \delta \big(z_{12} z_{34} \bz_{13} \bz_{24}-z_{13} z_{24} \bz_{12} \bz_{34} \big)  
					  \Theta\Big( \frac{  \eps_3 }{ \eps_1}\frac{   z_{23}  \bz_{34} }{z_{12}\bz_{14} } \Big)
				 	 \Theta \Big(  \frac{ \eps_3 }{ \eps_2} \frac{  z_{13} \bz_{34}  }{z_{12} \bz_{42}}\Big) 
					 \Theta \Big(  \frac{  \eps_3 }{ \eps_4}\frac{\bz_{13} z_{23}  }{\bz_{14} z_{42}  } \Big) .
		\end{split}
	\ee
	 Then, if we choose 
	 \be
	 	\alpha =  \Big(  \frac{  \eps_3 }{ \eps_4}\frac{\bz_{13} z_{23}  }{\bz_{14} z_{42}  } \Big)^{-1},
	 \ee
	 which is guaranteed to be positive by one of the step functions, we obtain a form of the amplitude in which we can easily study  the limit $\lambda_3 \to 0$.

	\subsection{ $n$-point MHV Yang-Mills amplitude }
	
The $n$-point MHV gluon amplitudes were derived in terms of generalized hypergeometric functions by Schreiber, Volovich and Zlotnikov  \cite{Schreiber:2017jsr}. In the appendix we verify  that these amplitudes are related by the conformally soft relation \eqref{MellinWard}.
 In this section, we present an alternate route to the same end. We first present a new formula for the $n$-point MHV celestial amplitude which manifestly exhibits all poles in $\lambda_i$ for $i = 3, ...,n$, allowing a streamlined  analysis of the  conformally soft limit.  Moreover, for $n >4$, the absence of poles in $\lambda_1, \lambda_2$ implies $\lim_{\lambda_i \rightarrow 0}\lambda_i \widetilde{\mathcal{A}}^{(n)}_{--+...+}(\lambda_i, z_i, \bz_i) = 0 \ {\rm for} \  i = 1,2$, 
 which is consistent with the vanishing of the momentum space soft limit  in these cases due to $\mathcal{A}^{(n-1)}_{-+...+}(\omega_i, z_i, \bz_i) = 0$. In the $n = 4$ case, the conformally soft limit of the negative helicity gluons in \eqref{Mellin4pointYang-Mills} yields the expected conformally soft formula.
 
The Mellin transform of the $n$-point MHV amplitude can be written
\be \label{nsoftlm}
\begin{split}
\widetilde{\mathcal{A}}^{(n)}_{--+...+}(\lambda_i, z_i, \bz_i)  &= \frac{1}{(-2)^{n-4}}\frac{z_{12}^3}{z_{23}\cdots z_{n1}} \prod_{i = 1}^n \left( \int_0^{\infty}d\omega_i \omega_i^{i\lambda_i} \right) \frac{\omega_1 \omega_2}{\omega_3 \cdots \omega_n} \delta^{(4)}\left(\sum_{i = 1}^n \epsilon_i \omega_i q_i\right)\\
&=\frac{1}{(-2)^{n-4}}\frac{z_{12}^3}{z_{23}\cdots z_{n1}}\prod_{i = 1}^{n} \left( \int_0^{\infty} d\omega_i \omega_i^{i\lambda_i - J_i} \right)\int \frac{d^4 y}{(2\pi)^4}e^{-i\sum\omega_i\left( \epsilon_i y \cdot q_i - i\varepsilon\right)} \\
&= \frac{1}{(-2)^{n-4}}\frac{1}{(2\pi)^4}\frac{z_{12}^3}{z_{23}\cdots z_{n1}}  \int d^4 y \prod_{i = 1}^n \frac{\Gamma(i\lambda_i + 1 - J_i)}{\left[i\epsilon_i y\cdot q_i +\varepsilon \right]^{i\lambda_i + 1 - J_i}}.
\end{split}
\ee
In the last line we used the formula
\be
\int_0^{\infty} e^{-a\omega} \omega^z d\omega = \frac{\Gamma(z + 1)}{a^{z+1}}.
\ee
We see explicitly from the  $\Gamma$ functions 
 that $\widetilde{\mathcal{A}}^{(n)}_{--+...+}(\lambda_i, z_i, \bz_i)$ has poles in $\lambda_i$ for $ 3 \leq i \leq n$ (but not in $\lambda_1$ or $\lambda_2$) so the conformally soft limit picks out the residue of the pole.  
 Since the factor $z^3_{12}/(z_{23} \cdots z_{n1})$ supplies the appropriate soft factor $z_{i-1~i+1}/(z_{i-1~i} z_{i~i+1})$, the amplitude obeys the conformally soft theorem so long as the remaining part becomes 
 independent of $(z_i, \bz_i)$ in the limit $\lambda_i \to 0$.  Indeed, in this limit, the factor associated to the $i$th particle  $\left[i\epsilon_i y\cdot q_i +\varepsilon \right]^{-i\lambda_i - 1 +J_i} \to 1$, so  the conformally soft formula  \eqref{MellinWard} holds.

\section*{Acknowledgements}
We are grateful to Nima Arkani-Hamed, Laura Donnay, Charles Marteau,  Andrea Puhm, Anders Schreiber, Marcus Spradlin, Tomasz Taylor and Ellis Yuan for useful discussions. This work was supported by DOE grant de-sc0007870. 

\begin{appendix}
\section{Schreiber-Volovich-Zlotnikov MHV amplitudes}	 

In this appendix we analyze the soft limits of MHV amplitudes using the representation of the 
$n$-point celestial amplitude in terms of generalized hypergeometric functions given by Schreiber, Volovich and Zlotnikov \cite{Schreiber:2017jsr}. They found\footnote{As in \cite{Schreiber:2017jsr}, we restrict our attention to a subset of the allowed configurations on the celestial sphere, for which the amplitudes can 
	be expressed in terms of generalized hypergeometric functions.}  
\be
\label{celMHV}
\begin{split}
\widetilde{\mathcal{A}}^{(n)}_{--+\cdots+}(\lambda_i, z_i, \bz_i)  = \frac{2\pi}{(-2)^{n-4}} \frac{z_{12}^3}{z_{23}...z_{n1}}\frac{1}{|U_{1,n}|}\delta(\sum_{j = 1}^n\lambda_j)\hat{\varphi}_n(\{\alpha\}, x)\prod_{i,j}\Theta\left(-\frac{U_{i,j}}{U}\right), 
\end{split}
\ee 
where $i = 1,...,n-4, j = n - 3,...,n$ and
\be 
\label{dhypgeo}
\begin{split}
&\hat{\varphi}_n(\{\alpha\}, x) = \frac{c_2}{c_1}\int_{u_1,..., u_3, 1 - \sum_{a = 1}^3 u_a \geq 0}\prod_{j = 1}^n P_j^{\alpha_j}\frac{dP_{n-3}\wedge dP_{n-2}\wedge dP_{n-1}}{P_{n-3}P_{n-2}P_{n-1}}, \\
& P_j = x_{0j} + x_{1j}u_1 + x_{2j}u_2 + x_{3j}u_3, \quad \quad   \frac{c_2}{c_1} = \frac{\Gamma(2 + i\lambda_1)\Gamma(2 + i\lambda_2)\prod_{j = 3}^{n-4}\Gamma(i\lambda_j)}{\Gamma(1 - i\lambda_1)\prod_{i = 1}^3\Gamma(1 - i\lambda_{n-i})}
\end{split}
\ee
are dual integral representations of hypergeometric functions. The parameters $\alpha_i$ and $x_{ij}$ are given by  
\be 
\label{ax}
\begin{split}
&\alpha_1 = 1,\  \alpha_2 = -2 - i\lambda_2,\  \alpha_3  = -i\lambda_3,..., \alpha_{n-4} = -i\lambda_{n-4},\  \alpha_{n-i} = -i\lambda_{n-i} +1,\  \alpha_n = -i\lambda_n, \\
& x_{0j} = \frac{U_{j,n}}{U_{1,n}}, \ x_{ij} = \frac{U_{j,n - 4+ i}}{U_{1,n-4 +i}} - \frac{U_{j,n}}{U_{1,n}}, \ x_{0 n} = -\frac{U}{U_{1,n}} = -x_{i n}, \ x_{01} = 1, \\
& x_{1 n-3} = -\frac{U}{U_{1,n - 3}}, \ x_{2 n-2} = -\frac{U}{U_{1,n-2}},\ x_{3 n-1} = -\frac{U}{U_{1,n-1}},\\ 
&{\rm for}\ i = 1,2,3, \ j = 2,...,n-4,
\end{split}
\ee
with
\be \label{Udef}
	\begin{split}
		U &= \eps_{n-3}\eps_{n-2}\eps_{n-1}\eps_{n}\det\left(q_{n-3} q_{n-2} q_{n-1} q_n\right) , \\
		 U_{i,j}& = \eps_{i} \eps_{j}\eps_{n-3}\eps_{n-2}\eps_{n-1}\eps_{n}\ \det\left(q_{n-3}...q_{j\rightarrow i}... \right).
	\end{split}
\ee

First, notice that the factor $c_2/c_1$ has poles in $\lambda_j$ for $j = 3, \cdots, n-4$, so the conformally soft limit of any of these gluons will be given by the
residue.  Since, as discussed previously, the factor $z^3_{12}/(z_{23} \cdots z_{n1})$ gives the appropriate soft factor, these amplitudes obey
the conformally soft theorem so long as $\hat \varphi_n$ becomes independent of $(z_j, \bz_j)$ in the limit $\lambda_j \to 0$.  The only dependence on $(z_j, \bz_j)$ in $\hat \varphi_n$ appears in $P_j^{-i \lambda_j}$, which behaves as 
$P_j^{-i \lambda_j} \to 1$ in the conformally soft limit.  Hence, 
$\hat \varphi_n$ becomes independent of $(z_j, \bz_j)$ and the celestial amplitude obeys the conformally soft theorem.

To obtain the conformally soft theorem for gluons $j = n-3, \cdots, n$, we work with an expression for the Mellin-transformed amplitude before the change of variables that gives \eqref{celMHV}
\be  \label{orig}
\widetilde{\mathcal{A}}_{--+\cdots +} =\frac{1}{(-2)^{n-4}} \frac{z_{12}^3}{z_{23}...z_{n1} |U|}\prod_{a = 1}^{n-4} \int_0^{\infty} d\omega_a \omega_a^{i\lambda_a}\frac{\omega_1 \omega_2}{\omega_3 \cdots \omega_{n-4}}
	\prod_{b = n-3}^n (\omega^*_b)^{i\lambda_b-1} \Theta(\omega_b^*),
\ee
where
\be
\label{sol-mc}
\omega_j^* = -\frac{1}{U}\sum_{i = 1}^{n-4} \omega_i U_{i,j}.
\ee
To take the conformally soft limit on the $n$th leg, we use the identity \eqref{deltaid1} so that the factor $(\omega^*_n)^{i\lambda_n-1}$ is turned into  $\delta(\omega_n^*)$.
As before, we need to show that the $n$-point amplitude becomes proportional to an $(n-1)$-point amplitude.  To this end, we can use  $\delta(\omega_n^*)$ to perform one of the $\omega_a$ integrals.   For example, we can solve $\omega_n^* = 0$ for $\omega_{n-4}$ 
\be
\label{newo}
\omega_{n-4} = -\sum_{i = 1}^{n-5} \frac{U_{i,n}}{U_{n-4,n}}\omega_i  ,
\ee
and perform the integral with respect to $\omega_{n-4}$.  We obtain an expression similar to  \eqref{orig}
\be  \label{appso}
\begin{split}
\lim_{\lambda_n \rightarrow 0}i\lambda_n \widetilde{\mathcal{A}}_{--+...+} =  \frac{1}{(-2)^{n-4}} \frac{z_{12}^3}{z_{23}...z_{n1} |U_{n-4, n}|}\prod_{a = 1}^{n-5} \int_0^{\infty} d\omega_a \omega_a^{i\lambda_a}
	\frac{\omega_1 \omega_2}{\omega_3\cdots \omega_{n-5}}\\
	\times \prod_{b = n-4}^{n-1} (\omega^*_b)^{i\lambda_b-1}\Theta(\omega_b^*)   ,
	\end{split}
\ee
where $\omega_{n-4}^*$ is given by the right-hand side of \eqref{newo} and
\be
\label{sol-nm1}
\omega_b^* = -\frac{1}{U}\sum_{i = 1}^{n-5} \left(U_{i, b} - \frac{U_{n-4,b} U_{i,n}}{U_{n-4,n}} \right)\omega_i, \qquad b = n-3, n-2, n-1.
\ee
To recognize that the right-hand side of \eqref{appso} is proportional to an $(n-1)$-point amplitude, we put it in the form \eqref{orig}.   
Introducing  analogues of \eqref{Udef} where $n-3, \cdots, n$ are replaced with $n-4, \cdots, n-1$
\be
	\begin{split}
		U'& =\eps_{n-4} \eps_{n-3} \eps_{n-2} \eps_{n-1}\det(q_{n-4} q_{n-3} q_{n-2} q_{n-1} ),\\
				 U'_{i,j}& = \eps_{i} \eps_{j}\eps_{n-4}\eps_{n-3}\eps_{n-2}\eps_{n-1}  \det\left(q_{n-4}...q_{j\rightarrow i}... \right),
	\end{split}
\ee
and using the identity
\be 
U_{i, b}U_{n-4, n} - U_{i,n}U_{n-4,b} = -U U'_{i,b},
\ee
we find \eqref{appso} can be written as 
\be  \label{appso1}
\begin{split}
\lim_{\lambda_n \rightarrow 0}i\lambda_n \widetilde{\mathcal{A}}_{--+...+} &= -\frac{1}{2}\frac{1}{(-2)^{n-5}}\frac{z_{n-1~1}}{z_{n-1~n} z_{n~1}} \frac{z_{12}^3}{z_{23}...z_{n-1~1} |U' |}\\
&\times \prod_{a = 1}^{n-5} \int_0^{\infty} d\omega_a \omega_a^{i\lambda_a}
	\frac{\omega_1 \omega_2}{\omega_3\cdots \omega_{n-5}} \prod_{b = n-4}^{n-1} (\omega^*_b)^{i\lambda_b-1}\Theta(\omega_b^*)   ,
	\end{split}
\ee
where now
\be
\label{sol-mc}
\omega_j^* = -\frac{1}{U'}\sum_{i = 1}^{n-5} \omega_i U'_{i,j},
\ee
so that right-hand side of \eqref{appso1} is readily identified as proportional to an $(n-1)$-point amplitude.
The conformally soft theorem of  \eqref{orig} for the gluons $j = n-3, n-2, n-1$ follows from similar arguments.

\end{appendix}

\providecommand{\href}[2]{#2}\begingroup\raggedright

\end{document}